\documentclass[sigconf]{acmart}
\renewcommand\footnotetextcopyrightpermission[1]{} 

\AtBeginDocument{%
  \providecommand\BibTeX{{%
    \normalfont B\kern-0.5em{\scshape i\kern-0.25em b}\kern-0.8em\TeX}}}

\setcopyright{acmcopyright}
\copyrightyear{2021}
\acmYear{2021}
\acmConference[SC '21]{The International Conference for High Performance Computing, Networking, Storage, and Analysis}{November 14--19, 2021}{St. Louis, MO, USA}
\acmBooktitle{The International Conference for High Performance Computing, Networking, Storage, and Analysis (SC '21), November 14--19, 2021, St. Louis, MO, USA}
\acmPrice{15.00}
\acmDOI{10.1145/3295500.3356219}
\acmISBN{978-1-4503-6229-0/19/11}

\usepackage{algorithmic}
\usepackage{graphicx}
\usepackage{textcomp}
\usepackage{xcolor}
\usepackage{comment}
\def\BibTeX{{\rm B\kern-.05em{\sc i\kern-.025em b}\kern-.08em
    T\kern-.1667em\lower.7ex\hbox{E}\kern-.125emX}}
\newcommand{\TODO}[1]{{\color{red}#1}}

\begin{document}

\title{Empirical Evaluation of  Circuit Approximations on Noisy Quantum Devices}
\thanks{
This work was supported in part by the Quantum Algorithm Teams program of the
Advanced Scientific Computing Research for Basic Energy Sciences
program, Office of Science of the U.S. Department of Energy under
Contract No. DE-AC02-05CH11231 and DE-AC02-06CH11357. It was also
funded in part by NSF awards DMR-1747426 and OAC-1917383.
This work used IBM Quantum services. The views expressed are those of
the authors, and do not reflect the official policy or position of IBM
or the IBM Quantum team.
}
\author{Ellis Wilson}
\affiliation{North Carolina State University}
\email{ejwilso2@ncsu.edu}

\author{Frank Mueller}
\affiliation{North Carolina State University}
\email{fmuelle@ncsu.edu}

\author{Lindsay Bassman}
\affiliation{Lawrence Berkeley National Lab}
\email{lbassman@lbl.gov}

\author{Costin Iancu}
\affiliation{Lawrence Berkeley National Lab}
\email{cciancu@lbl.gov}


\begin{abstract}
Noisy Intermediate-Scale Quantum (NISQ) devices fail to produce
outputs with sufficient fidelity for deep circuits with many gates
today. Such devices suffer from read-out, multi-qubit gate and
cross-talk noise combined with short decoherence times limiting
circuit depth. This work develops a methodology to generate shorter
circuits with fewer multi-qubit gates whose unitary transformations
approximate the original reference one. It explores the benefit of
such generated approximations under NISQ devices. Experimental results
with Grover’s algorithm, multiple-control Toffoli gates, and the
Transverse Field Ising Model show that such approximate circuits
produce higher fidelity results than longer, theoretically precise
circuits on NISQ devices, especially when the reference circuits have
many CNOT gates to begin with. With this ability to fine-tune
circuits, it is demonstrated that quantum computations can be
performed for more complex problems on today’s devices than was
feasible before, sometimes even with a gain in overall precision by up
to 60\%.

\end{abstract}

\acmVolume{}
\acmNumber{}
\acmArticle{}
\acmYear{2021}
\acmMonth{}
\acmISBN{}
\acmDOI{}

\maketitle
\pagestyle{plain} 

\section{Introduction}
 
Contemporary quantum computing devices are commonly referred to as
Noisy Intermediate-Scale Quantum (NISQ) computers as they are fraught by
a multitude of device, systemic, and environmental sources of noise that
adversely affect results of computations~\cite{preskill2018}. A
number of factors contribute to noise, or errors, experienced during
the execution of a quantum program. These include
\begin{itemize}
\item noise related to limits on qubit excitation time and program
  runtime due to decoherence;
\item noise related to operations, i.e., gates performing
  transformations on the states of one or more qubits;
\item noise related to interference from (crosstalk with) other
  qubits; and
\item noise subject to the process of measuring the state of a qubit
  via a detector when producing a program's output.
\end{itemize}
Efforts to reduce --- or otherwise mitigate --- noise are at the front
of the effort to create better, more practical quantum
computers today~\cite{Murali2019, murali2019noise, wilson20, scaffcc,
  preskill2018, ibm-pulses, openpulse,
  zulehner2018efficient,DBLP:conf/micro/TannuQ19,DBLP:conf/micro/TannuQ19a,Tannu:2019:QCE:3297858.3304007,li2019tackling,liquantum2020,wille2016look,siraichi2018qubit}.
Our work builds upon and complements these previous efforts.

All of these sources of noise have a common characteristic in that noise
becomes worse with circuit depth, i.e., the more sequential gates a
quantum circuit has, as quantum states (particularly excited states)
decohere over time. Today's NISQ devices feature qubits with
relatively short coherence times --- the longer an excited state has
to be maintained, the more noise is introduced, to the point where
eventually noise dominates and the original state become
unrecoverable. Current devices also suffer from noisy or imprecise
gates, which add imprecision to a circuit each time a gate is
applied, i.e., qubit state diverges slightly from the expected state
with the application of each transformation (rotation). Depending on the type of
gate, noise varies significantly: Gates operating on two qubits are an
order of magnitude more noisy than single qubit gates. Two qubit gates
also experience more cross-talk, due to interference with other qubits in close vicinity. Finally,
measurement of state (read-outs) is also subject to considerable
noise, on par with cross talk and two qubit gates, as opposed to said
single qubit noise.

A quantum program expressed as a circuit of gates operating on virtual
qubits needs to be translated into a sequence of pulses directed at
physical qubits. This translation step (a.k.a. transpilation) offers
optimization opportunities to reduce noise.  Besides translation of
pulses, quantum compilers consider secondary, noise-related objectives
to generate optimized quantum programs, e.g., by mapping virtual
qubits to less noisy physical qubits (in terms of readouts)~\cite{murali2019noise,Tannu:2019:QCE:3297858.3304007,DBLP:conf/micro/TannuQ19}
and their connections (for two qubit gates)~\cite{murali2019noise,DBLP:conf/micro/TannuQ19a}, or even by
increasing the distance to reduce cross-talk between qubits for a
given device layout~\cite{ibm-pulses,murali20,ding20}.

Another angle to address noise is to reduce the depth of circuits. By
reducing the number of gates in a circuit, especially the number of
two-qubit gates, the depth of the circuit, i.e., the span of time
during which qubits remain in excited states, is shortened, which
lowers the effect of decoherence. In fact, this may well bring long
circuits within reach of short decoherence times that otherwise could
not finish on a NISQ device before losing their states.
One promising way to reduce the number of gates is to create an {\em
  approximate circuit}~\cite{Amy2013,DeVos2015,alam20}, i.e.,
a circuit which does not provide an exact
(theoretically perfect) transformation for a target unitary but rather
a ``close'' fit for the unitary. (One could make a comparison to fixed
precision arithmetic in classical computation here, which often relies
on converging calculations as an approximation of exact numerical
results.) On a NISQ device, an exact quantum circuit is prone to develop
large error with increasing circuit depth. In contrast, a
near-equivalent approximate quantum circuit with shorter depth, even
though subject to a slightly incorrect transformation, may have the
potential to yield a result that is {\em closer} to the noise-free
(theoretically) desired output. This opens up an interesting trade-off
between longer-depth theoretical precision with more noise
vs. shorter-depth approximation with less noise. It is this trade-off
this work aims to assess and quantify.

The task of finding an approximate circuit is similar to the process
of circuit synthesis~\cite{qs, qfast}. Circuit synthesis is another
avenue that attempts to reduce circuit depth. Synthesis here refers to
the process of what can be considered design space exploration: Given
a quantum program, expressed as an exact circuit or an equivalent
unitary matrix, other circuits are systematically constructed and then
evaluated in a search for an equivalent, shorter depth quantum circuit
with the {\em same} unitary. If found, such a target circuit can be
transpiled to a specific machine layout and set of gates with shorter
execution time, which may be within the given decoherence threshold of
a NISQ device.

The main difference between circuit synthesis and searching for an
approximate circuit is that instead of searching for an equivalent
(functionally indistinguishable) circuit, the latter searches for an
{\em approximate} circuit of a shorter depth with a {\em slightly
  different unitary}. While this leads to inferior results on a
noise-free machine, the intuition is that due to noisy gates, shorter,
approximate circuits have the potential to outperform longer, more
precise circuits.

There are many different metrics which can be used to determine
whether two circuits are equivalent. Quantum synthesis
compilers~\cite{qs, qfast} typically use distance metrics between
``process'' representations of the program, such as the
Hilbert-Schmidt (HS) distance between the associated unitary matrices, or
the diamond norm~\cite{Gilchrist_2005, ahar97}.  In the process view, two programs are
deemed equivalent when at distance ``zero''.

In the context of this work aiming at approximation, synthesis is
used to find a circuit exceeding a distance of zero relative to the
original program so that, when run on a NISQ machine, its output is
expected to be close to that of the original program. One challenge
with using approximate circuits is that of finding a suitable metric
to assess the appropriateness of a set of approximate circuits. One
potential option is a process distance, such as HS, within a certain
range (threshold). Another is to instead consider output-related
metrics, such as the Jensen-Shannon Divergence or Total Variation
Distance~\cite{endres_03}. This remains an open question.

The novelty of this work is in its focus on the analysis of a
particular use-case of approximate circuits, namely by considering a
set of approximate circuits created by quantum synthesis software. 
When these offered an unworkable number of circuits, we constrained
which ones we used by a given HS distance as a threshold. We never 
choose an HS threshold of less than 0.1, which still results in a wide
range of approximate circuits. With a large selection of circuits we can
investigate the behavior of {\em many} approximate circuits in the
presence of {\em different noise levels}.

With this work, we make the following novel contributions to the
broader aim of searching for approximate circuits:

\begin{itemize}

\item We demonstrate how to obtain a wide range of approximate
  circuits from custom modified circuit synthesis tools.

\item We provide a proof-of-concept that approximate circuits can
  outperform exact circuits on NISQ devices for small, well known
  algorithms --- Grover's Algorithm and the Multiple-control Toffoli gate --- as well as
  a specific physics application, namely 
  the three-four qubit Transverse Field Ising Model (TFIM).

\item We show how the results of approximate circuits change relative
  to the noise induced by two-qubit errors. Specifically, we assess
  the effect of two-qubit errors of lower-depth circuits with
  different approximation thresholds vs. that of the exact, longer
  circuit. Experiments indicate improvements in overall precision for
  shorter approximate circuits over longer precise ones by up to 60\%.

\end{itemize}

\section{Problem Statement  and Objectives} %

This work seeks to assess if approximate circuits can outperform exact
circuits on today's NISQ devices.  Utilizing approximate circuits
ultimately comes with challenges posing four fundamental questions:
\begin{enumerate}
\item How can approximate circuits be generated?
\item Can the search for or generation process of approximate circuits
  be constrained and, if so, how?
\item Will the resulting approximate circuits outperform their
  equivalent original ones?
\item Can algorithms be designed to make circuit synthesis and
  the search of resulting circuits scalable?
\end{enumerate}

Before investigating these problems, however, a more fundamental
question should be asked: {\em Is there any value in approximate
  circuits to begin with?} In other words, can any approximate circuit
actually outperform the original circuit at all?  It is {\em this}
line of reasoning that our work is trying to answer --- before we can
explore the more general challenges posed by the four questions above.

In this work, we show that there {\em is} potential value in
approximate circuits: They can outperform theoretically perfect
circuits on today's NISQ hardware. We also confirm that any method of
selecting appropriate approximate circuits will need to take the
noise/error levels of target devices into account.

\section{Design}

\begin{figure*}[ht!]
     \centering
     \includegraphics[width=\textwidth]{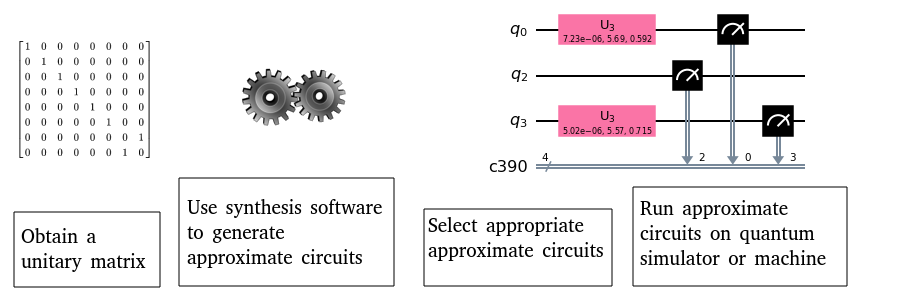}
\vspace*{-\baselineskip}
     \protect\caption{Generic workflow of using approximate circuits. The example is an approximation of the first timestep of the TFIM circuit.}
\label{workflow}
\end{figure*}

One way to find approximate circuits is to look at approximate 
circuits generated by the intermediate steps of circuit 
synthesis programs. These programs do not typically scale to a
level which would make them an ideal way to create large approximate
circuits in practice, but are an easy way to create them as a
proof of concept. Synthesis programs typically look for the 
shortest circuit they can find with a distance of some kind at 
``zero''. As these programs are interested in finding 
as short a circuit as possible, they tend to investigate many 
shorter circuits before finding their target; these circuits 
are already nearly optimized for 
their layout, making them ideal approximate candidates.

Before utilizing synthesis software to generate approximate
  circuits, we typically need to alter the synthesis tools to produce
  as output, besides a single circuit, additional circuits that are
  farther away from the target. These tools already generate and test
  many circuits during the search for an equivalent circuit, which
  allows our enhancements to integrate naturally with the existing
  flow within synthesis tools.
%
%

Figure~\ref{workflow} shows the workflow of our process. We first
need to obtain our target unitary.
Quantum operations can be represented by matrices, and the
  target unitary is the result of multiplying these transforming
  matrices of a circuit (or subcircuit) that is to be approximated. In
  IBM's Qiskit python interface~\cite{Qiskit}, the unitary of a
  circuit can be obtained
with the following command on the target QuantumCircuit object
$circuit$:

$$ matrix = qiskit.quantum\_info.Operator(circuit).data $$

The second step is to use our altered synthesis software to
  generate approximate circuits for our target matrix.

Third, given the the enhanced synthesis software that outputs every
circuit it checks, we need to select which approximate circuits we
want to check. How to perform this selection is still an open
question. For our analysis, we intend to compare a large number of
circuits, so we select many circuits with little to no filter for
pre-selecting the most accurate ones.

Finally, the selected circuits need to be run on a quantum
machine or simulator. For our study, we then compare the 
results with the expected output, either a known value or
our original circuit run on a simulator with no errors.

\section{Implementation}

For our exploration of approximate circuits, we use two different 
synthesis tools, QSearch and QFast, both of which are part of the 
Berkeley Quantum Synthesis Toolkit (BQSkit).

The QSearch~\cite{qs} optimal depth circuit synthesis tool
builds a sequence of circuits of increasing length and decreasing
HS distance until it finds the first circuit with a distance of
``zero'', a value which can be specified but which defaults to less
than 1e-10. It does this by following the A* algorithm. Specifically, 
it explores different branches of the circuit space by adding on 
blocks of three gates. Certain machine layouts can be taken into 
account by restricting these blocks to only being placed between 
connected qubits. These blocks are made up of one two qubit 
controlled NOT (CNOT) gate and two single qubit U3 gates on each of 
the same qubits.  The U3 parameters are optimized using one of a 
number of different numerical optimizers, including COBYLA and BFGS,
provided by SciPy 1.20, and reoptimized after each step. This 
optimization ensures that, for this specific layout of CNOT and U3 
gates, this circuit is the closest possible to the target. Because 
it considers each option, this is guaranteed to be depth optimal 
with respect to two-qubit gates. 

The QFast~\cite{qfast} synthesis tool likewise builds a sequence of
circuits of increasing length, but it has a more complicated algorithm
for finding circuits of increasingly higher quality. QFast is not
guaranteed to be optimal and gives less of a choice of approximate
circuits, but handles circuits with more qubits than QSearch within
acceptable search times.

In our work we enhance the QSearch software such that instead of
saving only the final circuit, it also saves every intermediate
circuit during its search. We then select a portion of the circuits,
always with a maximum HS distance threshold of at least 0.1, in order
to have a wide range of circuits but none which differ entirely from
the target circuit. QFast requires no source code alteration, but it
needs to be given a dictionary with the key of
``partial\_solution\_callback'' pointing to a function to output these
solutions. This dictionary is used by calling QFast with the keyword
``model\_options''.

These circuits are then executed in three different methods. First, they are
executed on the IBM
Qiskit~\cite{ibm-noise-sim} simulator using hardware specific
(ibmq\_ourense, ibmq\_toronto, ibmq\_manhattan, ibmq\_rome, ibmq\_santiago) noise models. These noise models are created 
using error data collected from IBM's own physical machines, creating a
noisy simulator. 

Second, they are executed on noise level sweeps. These use the ibmq\_ourense
noise model as a base, but change the two-qubit gate noise level
during a sensitivity study in order to observe the effect of different
types and levels of noise.

Finally, The approximate circuits are also executed on the ibmq\_ manhattan, ibmq\_toronto, and ibmq\_rome
physical machines.

\section{Experimental Framework}

We selected three different algorithms for the evaluation. 
We start with circuits
generated~\cite{bassman2020towards, bassman2021constantdepth} for the time-dependent Transverse
Field Ising Model (TFIM). The TFIM is a quintessential model for
studying various condensed matter systems, and its time-dependent
manifestation shows promise for revealing new information about
non-equilibrium effects in materials.  Current algorithms for
designing quantum circuits for the simulation of such models, however,
produce circuits that increase in depth with the growing number of time-steps;
circuits quickly grow beyond the NISQ fidelity budget, placing
tight limits on the number of time-steps that can be simulated.
This class of circuits, therefore, stands to greatly benefit from
shorter, approximate circuits. In addition, the output for these
circuits can be condensed to a single number to easily be compared 
to the output of the approximate circuits, allowing for an easy
target for the approximate circuits.

We next study Grover's algorithm~\cite{Grover1996:search} followed by
the multi-control Toffoli gate~\cite{Toffoli_1980} to demonstrate the
general capability of our method.

We decided to focus on small circuits for this work due to NISQ and
synthesis limitations. We use the three and four qubit execution of
the circuits for most of our experiments, and scale up to five qubits
with the multi-control Toffoli gate. For TFIM, we assess at the first
21 time steps of 3ns. This results in 21 different circuits for
different times in the evolution of the magnetization. All of these
circuits are related, but they can also be investigated individually.

\begin{table}[ht!]
\centering
\begin{tabular}{||c|c|c||} 
\hline
IBM Machine & Num. qubits & Av. CNOT err. \\ [0.5ex] 
 \hline\hline
 Manhattan & 65 & .01578 \\ 
 \hline
 Toronto & 27 & .01377 \\
 \hline
 Santiago & 5 & .01131 \\
 \hline
 Rome & 5 & .02965 \\
 \hline
 Ourense & 5 & .00767 \\
 \hline
\end{tabular}
\protect\caption{Average CNOT errors on a selection of
IBM physical machines as of 2021/01/18}
\label{error_table}
\end{table}

Table~\ref{error_table} provides a snapshot of typical CNOT error rates at 
the time of writing. they give a contemporary view of the types of CNOT 
errors that we compare against and reflect the constant changes of NISQ 
devices with different error rates on different qubit connections even on 
the same device.

For our experiments using simulators we transpile under IBM's
optimization level 1 with mappings to qubits 0, 1, 2, 3, and 4.  Our
experiments on physical machines are transpiled under optimization
level 3, which at the time of writing allows IBM to map virtual qubits
to the best available physical qubits. All work is performed with
Python 3.8.2 and Qiskit 0.18.3, Qiskit-aer 0.5.1, Qiskit-ibmq-provider
0.6.1, and Qiskit-terra 0.13.0. Our QSearch enhancements are based on
search\_compiler version 1.2.1, and we used QFast version 2.1.0.

\section{Results}

We first report experimental results for simulations under given noise
models of contemporary quantum devices subject to NISQ constraints. We
then perform a sensitivity study on the effect of noise levels,
including both smaller (future) and larger (past) noise levels than
seen on the reference device, still using simulation. This is followed
by experiments on IBM Q devices with approximate circuits under
default transpilation with full optimization. Finally, we perform a
sensitivity study investigating the effect of how approximate circuits
are mapped to qubits on hardware devices with respect to noise level,
particularly of CNOT gates.

\subsection{Noise Model Simulations}

\begin{figure}[htb]
  \centerline{\includegraphics[width=\columnwidth]{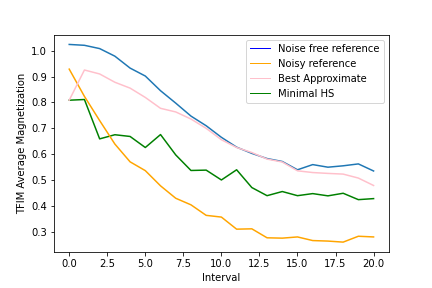}}
  \vspace*{-\baselineskip} \protect\caption{Magnetization over 21
    timesteps of selected (best/minimal HS) approximate circuits for
    the 3-qubit TFIM using the Toronto error model.}
\label{tor_lines}
\end{figure}

We first investigate the noise and approximation quality of our approach.
%
Figure~\ref{tor_lines} depicts results for a 3-qubit TFIM problem
under the Toronto (IBM Q) noise model with magnetization (y-axis)
over time steps (x-axis) in 21 intervals of 3ns
each.
Series ``Noise free reference'' shows the result for the circuit
generated by the TFIM domain generator and simulated on the ideal
hardware. This is the target for the other circuits; the closer they are to these results, the better they are. Series ``Noisy reference'' shows the behavior of the same
circuits when simulated with the hardware specific noise model.
``Noisy reference'' behavior quickly diverges from the ideal as
circuits become more complex with increasing timesteps. Series
``Minimal HS'' shows the behavior of the synthesized circuits when
using process metrics (HS) as the quality indicator. As these are much
shorter (six CNOTs versus tens of CNOTs for the reference circuit)
than the baseline implementation, their results are typically closer to
the ideal results.

The potential of approximate circuits is depicted by Series ``Best
approximate'', where we select the circuits with ``best'' output
behavior. Their CNOT depth is always shorter than the HS=0 circuits,
and so even though the process distance is greater they provide a
result closer to the noise free reference. This was also observed
across other noise models.

\textbf{Observation 1: Short approximate circuits can outperform long
  circuits with a lower process distance in simulation under device
  noise models.}

\begin{figure}[htb]
  \centerline{\includegraphics[width=\columnwidth]{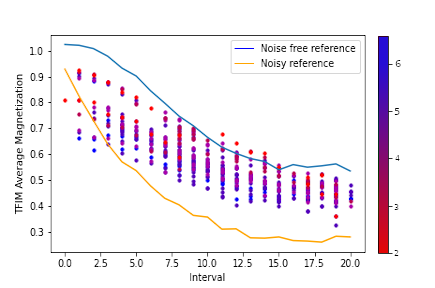}}
  \vspace*{-\baselineskip} \protect\caption{Magnetization over 21
    timesteps of all approximate circuits for the 3-qubit TFIM
    using the Toronto error model.}
\label{tor_dots}
\end{figure}

Let us investigate the range of solutions generated by approximate
circuits in more detail.
Figure ~\ref{tor_dots} shares the noise free and noisy reference data
series with Figure~\ref{tor_lines}, but it additionally includes dots
representing each approximate circuit.  The colors of the dots
indicate how many CNOTs were used in the approximate circuits; in this
case, red dots represent two CNOTs and blue dots represent six.  It
can be seen that while there was a wide difference in accuracy over
the different approximate circuits, nearly all of them performed
better than the noisy reference.

\begin{figure}[htb]
  \centerline{\includegraphics[width=\columnwidth]{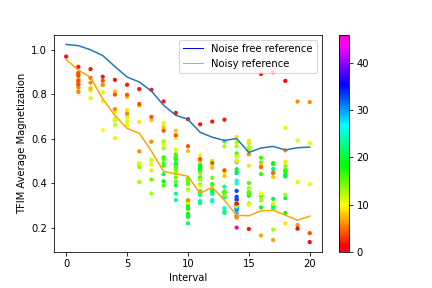}}
  \vspace*{-\baselineskip} \protect\caption{Magnetization over 21
    timesteps of approximate circuits for 4 qubit TFIM sing the
    Santiago noise model.}
\label{TFIM4}
\end{figure}

We next investigate the impact of circuit width (in qubits) and depth
(in CNOT gates) for the same TFIM application.
Figure~\ref{TFIM4} represents the four qubit TFIM circuit with the
same line graphs again. The number of CNOTs in an individual circuit
in this case can range from 1 to 48, which illustrates the wide range
of approximate circuits, many of which are closer to the noiseless
reference than the noisy reference is.

\begin{figure}[htb]
  \centerline{\includegraphics[width=\columnwidth]{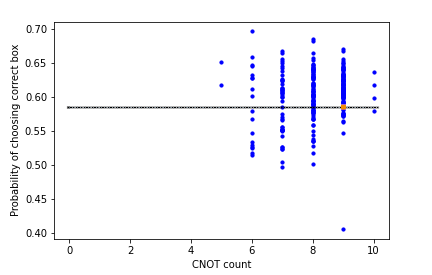}}
  \vspace*{-\baselineskip} \protect\caption{Probability of correct
    result over CNOT count of approximate circuits for 3 qubit
    Grover's algorithm using the Toronto noise model. Reference
    circuit in red.}
\label{grover_dots}
\end{figure}

We now turn our investigation to the impact of circuit approximation
for different algorithms and circuits, first with Grover and then with
Toffoli.
Figure~\ref{grover_dots} depicts results for Grover's algorithm with a
search target of '111' over eight boxes, where each dot represents a
circuit. The blue dots each indicate an approximate circuit, while the
orange dot and the line represent the output circuit of the hand-coded
reference implementation with nine CNOTs.  Figure~\ref{grover_dots}
shows the quality of the circuits as the probability of selecting the
correct box (y-axis), where higher probability is better.  Here, CNOT count is shown on
the x-axis rather than indicated by color coding. This shows a wide
array of approximate circuits, many of which outperform the reference;
only a smaller fraction (below the dashed line) underperform. The
challenge here is to select a ``good'' approximate circuit from the
wide array of possible candidates. We observe and investigate this
challenge with different metrics but its solution is ultimately beyond
the scope of this paper.

\textbf{Observation 2: To capitalize on the potential of approximate
  circuits, a selection method and an associate metric are required to
  ensure superior performance under noise.}

\begin{figure}[htb]
  \centerline{\includegraphics[width=\columnwidth]{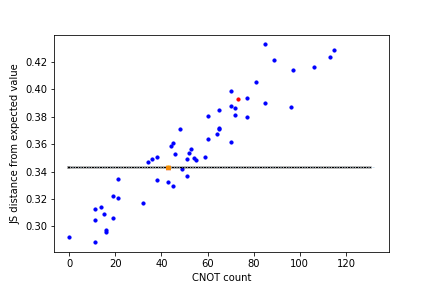}}
  \vspace*{-\baselineskip} \protect\caption{Jensen Shannon (JS)
    distance over CNOT count of approximate circuits for 4 qubit
    Toffoli compared to the reference circuit using the Manhattan
    noise model. Qiskit (orange) and QFast (red) circuits are
    outperformed by other approximate circuits.}
\label{toffoli4}
\end{figure}

We further perform experiments for the Toffoli gate with different
numbers of qubits.
Figure~\ref{toffoli4} shows the results for four qubit Toffoli
gate\textemdash that is, three control qubits to one target qubit. We
use the Jensen Shannon (JS) distance\cite{endres_03} to analyze these
circuits (y-axis), as the Toffoli gate can be programmed to
represent a variety of functions, each with different (but known)
output.  We test each approximate circuits for a subset of such
functions and parameters since a given circuit results in different
probabilities for correct output. The JS distance provides a composite
metric to reflect accuracy (lower is better in this case).

The four qubit results indicate that low-depth approximate circuits
outperform those with high CNOT depth. The orange dot on the dashed
line represents Qiskit's multiple-control Toffoli gate without any
ancilla bits while the red dot indicates QFast's default result of
an equivalent circuit.
The JS metric indicates that the former (orange) outperforms
the latter (red). Furthermore, many deeper approximate circuits
perform worse than Qiskit's Toffoli without ancilla while shorter
approximations (below the line) can provide even better results than
Qiskit. This implies that there is room for improvement even over the
reference implementation of given circuits on today's noisy machines.

\textbf{Observation 3: Approximate circuits generated from synthesis
  can outperform discrete reference circuits under noise.}

\begin{figure}[htb]
  \centerline{\includegraphics[width=\columnwidth]{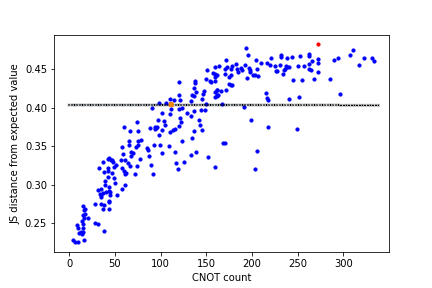}}
  \vspace*{-\baselineskip} \protect\caption{JS distance over CNOT
    count of approximate circuits for 5 qubit Toffoli compared to the
    reference circuit using the Manhattan noise model.}
\label{toffoli5}
\end{figure}

Figure~\ref{toffoli5} depicts results for a five qubit Toffoli gate,
again without any ancilla qubits for either the reference or
approximate circuits. These results reinforce the earlier four qubit
results: The JS distance of the reference circuit is higher for five
qubits, but some approximate circuits have a distance even closer to
zero than the best of those for the four qubit Toffoli gate. The
correlation of shorter circuits performing better is evident, but
outliers exist. As the number of CNOTs increases to the hundreds, the
JS values approach 0.465. This is significant because, in this
implementation, random noise (an equal number of results of 00000 as
00001 as 00010 and so on) results in a JS distance from the target of
0.465.

We also performed experiments for a 3-qubit Toffoli gate. In this
case, the 3-qubit approximate circuits performed poorly compared to
the optimized hand-crafted Toffoli gate commonly used, which uses only
6 CNOTs (graph omitted). This illustrates that simple, short circuits
provide little benefit for approximations via QSearch or QFast whereas
deeper and more complex ones can benefit significantly for today's
noisy quantum hardware. It also presents a challenge for synthesis
tools as wider circuits (beyond 6-8 qubits) with corresponding depth
results in excessive search cost.

\textbf{Observation 4: The benefit of using approximate circuits
  increases with the depth of the reference circuit.}

\subsection{Error Sensitivity Studies}

\begin{figure}[htb]
  \centerline{\includegraphics[width=\columnwidth]{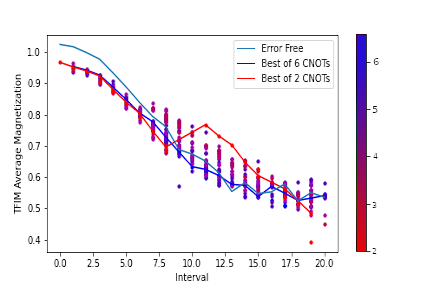}}
  \vspace*{-\baselineskip} \protect\caption{Magnetization over 21
    timesteps of approximate circuits for 3-qubit TFIM using the
    Ourense noise model with no CNOT error.}
\label{red1}
\end{figure}

\begin{figure}[htb]
  \centerline{\includegraphics[width=\columnwidth]{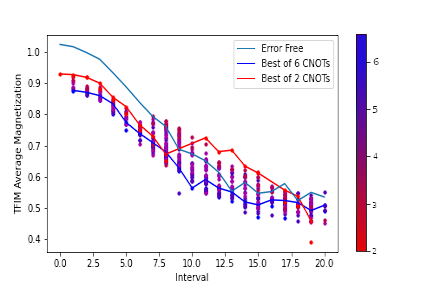}}
  \vspace*{-\baselineskip} \protect\caption{Magnetization over 21
    timesteps of approximate circuits for 3-qubit TFIM using the
    Ourense noise model with a simulated CNOT error of 0.12.}
\label{red2}
\end{figure}

\begin{figure}[htb]
  \centerline{\includegraphics[width=\columnwidth]{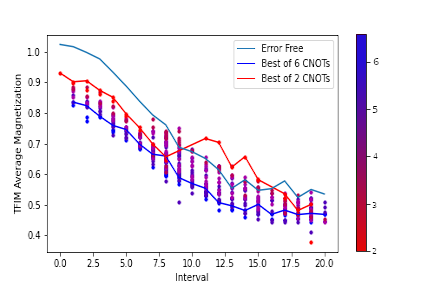}}
  \vspace*{-\baselineskip} \protect\caption{Magnetization over 21
    timesteps of approximate circuits for 3-qubit TFIM using the
    Ourense noise model with a simulated CNOT error of 0.24.}
\label{red3}
\end{figure}

We assess the sensitivity of approximate circuits to noise. To
this end, we use the Ourense noise model as a base but change the
CNOT error rate to assess how the performance of circuits changes
results in response.  Figures~\ref{red1},~\ref{red2}, and~\ref{red3}
present
the approximate circuits with increasing noise. Circuits are again color
coded using their depth, with red circuits consisting of two CNOTs and blue
circuits of six. The lines of these colors represent the best
performing approximate circuits for that number of CNOT gates.

Figure~\ref{red1} depicts simulations for a CNOT noise level of
zero. It illustrates the spread of circuits with different noise
sources (with the exception of CNOT noise), and shows that CNOT depth is
not closely correlated to the quality of results with no CNOT noise.

Figure~\ref{red2} shows the simulation for a CNOT error of 0.12,
similar to that of today's lowest quality physical devices, and
assesses the impact on performance.  Note that the increase in CNOT
error is accompanied by a decrease in the observed average
magnetization. Many of the longer circuits in blue or purple, which
were covered up by the red dots, become visible showing that a diverse
number of approximate circuits react differently under CNOT noise.

Figure~\ref{red3} depicts simulations for a CNOT error of 0.24, which
is worse than many current IBM machines and reinforces this trend. These
results are promising. We clearly see that some individual circuits
improve, i.e., more closely approximate the error free reference, with
an increase in two-qubit error. We also see that deeper circuits are more affected by CNOT error than
shallower circuits. With a low two-qubit error, many of the deeper
circuits lie on the line corresponding to the error free reference. As
this error increases, these deep circuits quickly decline in quality, and the shallower circuits perform relatively better. This
is seen as the best of the longest circuits perform worse than the
best of the shortest circuits for all timesteps; but without CNOT
noise, this is not necessarily true.  Some of these circuits actually
benefit from the noise and more closely approximate the error free
reference.

The takeaway from this trend is that different approximate circuits
should be chosen based on the error levels of the physical machine. A
program can afford to use a long circuit on machines with low error,
but a noisier machine will benefit from a shorter, approximate
circuit.

\textbf{Observation 5: Beyond merely being less affected by noise than the reference circuit, some approximate circuits perform better in
  the presence of noise. This performance increase is dependent on the noise parameters of the system.}

\begin{figure}[htb]
  \centerline{\includegraphics[width=\columnwidth]{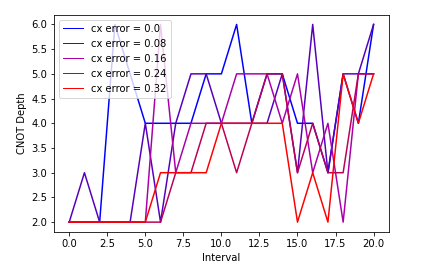}}
  \vspace*{-\baselineskip} \protect\caption{CNOT depth over 21
    timesteps of approximate circuits for TFIM showing the best
    approximate circuits for select CNOT errors.}
\label{simulateallsweepdepths}
\end{figure}

Figure~\ref{simulateallsweepdepths} further supports this by 
depicting the depth of the best performing circuit for different
noise levels. A trend can be seen: the worse the error (the more red the line),
the shallower the circuits with the highest performance in general,
but not under all circumstances.  A similar trend is seen with our
other algorithms (figures omitted).

These results generally support the initial conjecture: as the amount
of noise in the models increases, the output quality of deeper
circuits deteriorates more quickly than that of the shallower circuits. This
causes some of the shallower circuits to produce results that are
closer to the ideal results than the deeper circuits, even though the
deeper circuits would perform better on an ideal, noise-free
machine. This is most noticeable with circuits that contain many CNOTs
and on noisier models. It is less noticeable with circuits which are
already short or simulated on models of low noise.

\textbf{Observation 6: The greater the level of two-qubit noise on the
  target machine, the more benefit is gained from short approximate
  circuits.}

\subsection{Results on IBM Q Hardware}

\begin{figure}[htb]
  \centerline{\includegraphics[width=\columnwidth]{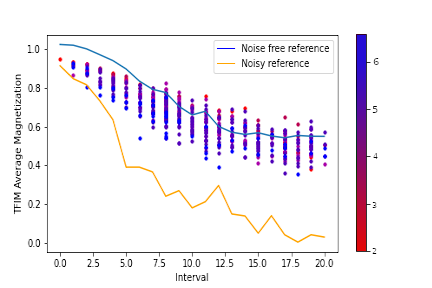}}
  \vspace*{-\baselineskip} \protect\caption{Magnetization over 21
    intervals of approximate circuits for 3 qubit TFIM on the
    Manhattan physical machine.}
\label{TFIM3_real}
\end{figure}

\begin{figure}[htb]
  \centerline{\includegraphics[width=\columnwidth]{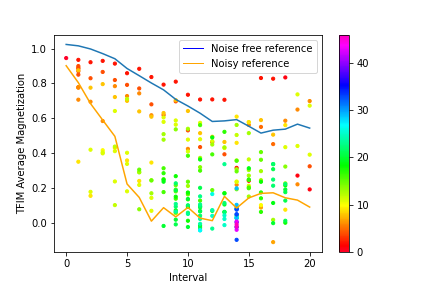}}
  \vspace*{-\baselineskip} \protect\caption{Magnetization over 21
    timesteps of approximate circuits for 4 qubit TFIM on the
    Manhattan physical machine.}
\label{TFIM4_real}
\end{figure}

Figures~\ref{TFIM3_real} and~\ref{TFIM4_real} depict results from
running the three and four qubit TFIM circuits on contemporary IBM
quantum hardware devices.  These results provide insight on how much
can be gained from using approximate circuits in practice today. We
observe that almost all of the approximate circuits in
Figure~\ref{TFIM3_real} and the large majority of the approximate
circuits in Figure~\ref{TFIM4_real} perform better than the default
circuits.


We also observe that the approximate circuits here are distributed
similarly to Figure~\ref{red2}, showing that the earlier constructed
noise models are not far off from actual noise on hardware today.

\textbf{Observation 7: Approximate circuits can perform well compared
  to reference circuits on real quantum hardware devices as well as on
  noisy simulators.}

\begin{figure}[htb]
  \centerline{\includegraphics[width=\columnwidth]{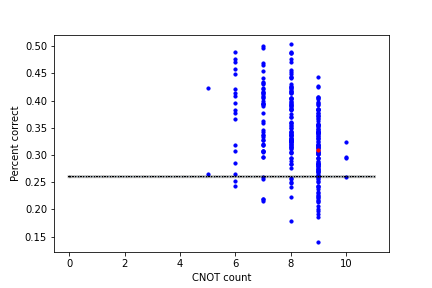}}
  \vspace*{-\baselineskip} \protect\caption{Probability of correct
    results over CNOT count of approximate circuits for 3 qubit
    Grover's Algorithm on the Rome physical machine.}
\label{grover_real}
\end{figure}

Figure~\ref{grover_real}, similar to Figure~\ref{grover_dots}, depicts
results from experiments with the 3 qubit implementation of Grover's
algorithm. As before, many (but not all) of the approximate circuits
perform better than the reference circuit. There is a minor bias to
shorter circuits performing better, but not a significant one.
It should be noted that the reference circuit here had more
  than 50 CNOTs and is thus omitted from the figure. The line is
  still at the performance level of the reference circuit.

\begin{figure}[htb]
     \centerline{\includegraphics[width=\columnwidth]{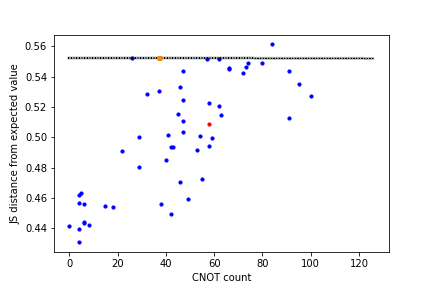}}
\vspace*{-\baselineskip}
     \protect\caption{JS distance over CNOT count of approximate
       circuits for 4 qubit Toffoli on the Manhattan physical machine.}
\label{toffoli4_real}
\end{figure}

Figure~\ref{toffoli4_real} shows the result of the 4 qubit Toffoli and
its approximates on a real machine.  At first, the result looks
similar to the distribution in Figure~\ref{toffoli4}. However, while
the best approximate circuits do have a much lower JS score (by 78\%)
than the reference circuit (orange), the reference circuit and many of the
approximate circuits actually perform worse than random noise (as
mentioned in the context of discussing Figure~\ref{toffoli5}, random noise has a distance of 0.465).
This indicates that even the approximate circuits are still too noisy
to run on the physical machines, but we expect them to perform better
than the reference circuit when run on less noisy devices.

\textbf{Observation 8: Trends indicate a continuing potential of
  approximate circuits to outperform reference circuits in the near
  future, even as noise levels in physical machines decline.}

\subsection{Sensitivity to Qubit Mappings on IBM Q Hardware}
\begin{figure*}[t]
     \centering
     \includegraphics[width=\textwidth]{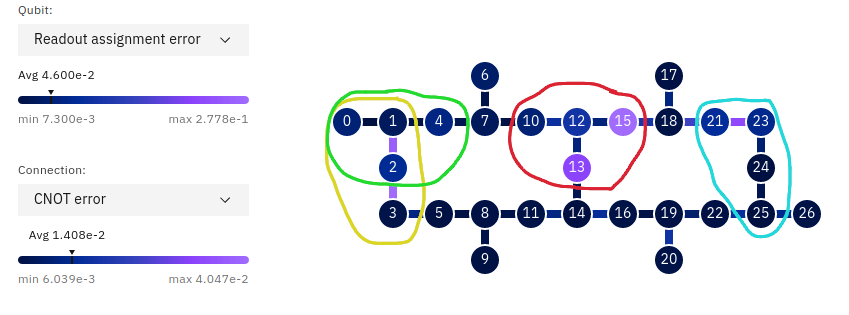}
\vspace*{-\baselineskip}
     \protect\caption{Noise report from IBM for their Toronto machine
       at the time of study. Different circles represent
       different mappings.}
\label{mapping}
\end{figure*}

We further investigate the impact of mapping circuits to specific
qubits with CNOT resonance channels of different noise levels for the
IBM Toronto physical quantum device using the 4 qubit Toffoli. The
qubit connectivity of this machine is shown in Figure~\ref{mapping},
as reported by IBM on the day of experimentation.  The nodes represent
different qubits, and their color indicates the readout error in the
range depicted on the upper heatmap index to the left.  The edges
represent the connection between the qubits, and their color indicates
the CNOT error level on the lower heatmap index.

Experiments are conducted with four different (manual) mappings for
the approximate circuits plus one (automatic) mapping using Qiskit's
transpiler at optimization level 3. We depict only the circuits with
the best and worst results here.

\begin{figure}[htb]
  \centerline{\includegraphics[width=\columnwidth]{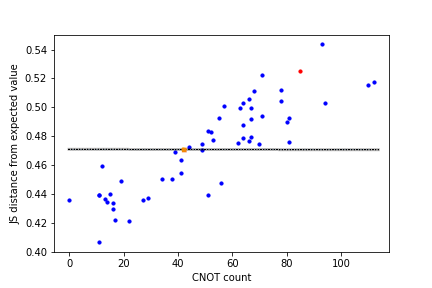}}
  \vspace*{-\baselineskip} \protect\caption{JS distance over CNOT
    count of approximate circuits for 4 qubit Toffoli on the Toronto
    physical machine showing the best performing mapping.}
\label{toffoli4_best}
\end{figure}

Figure~\ref{toffoli4_best} shows results for circuits mapped onto the
qubits within the blue circle in Figure~\ref{mapping}.  These results
exhibit the shortest JS distance of $\approx$0.4 (best), and about a
third of the circuits lie below the reference of $\approx$0.47.

\begin{figure}[htb]
  \centerline{\includegraphics[width=\columnwidth]{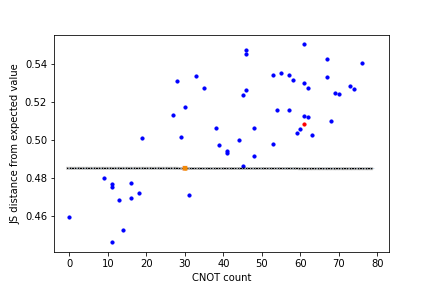}}
  \vspace*{-\baselineskip} \protect\caption{JS distance over CNOT
    count of approximate circuits for 4 qubit Toffoli on the Toronto
    physical machine showing the worst performing mapping.}
\label{toffoli4_worst}
\end{figure}

Figure~\ref{toffoli4_worst} depicts the results for mappings into the
red circle, which provided the worst results with higher JS distance
(reference: JS$\approx$ 0.485, approximate circuits start at
JS$\approx$0.45) than that of any other mapping. Other mappings (not
depicted) lie in between these results.

\begin{figure}[htb]
     \centerline{\includegraphics[width=\columnwidth]{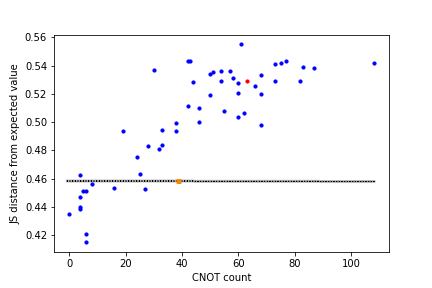}}
\vspace*{-\baselineskip}
     \protect\caption{JS distance over CNOT count of approximate
       circuits for 4 qubit Toffoli on the Toronto physical machine
       showing mappings generated by Qiskit with optimization level 3.}
\label{toffoli4_ol3}
\end{figure}

Figure~\ref{toffoli4_ol3} shows the results of transpiling the same
approximate circuits with Qiskit under level three optimizations. As
each approximate circuit was mapped individually and automatically by
Qiskit, no single mapping can be reported. The green circle shows the
mapping for the best performing circuit within that run, and yellow
indicates that of the reference circuit. Fewer circuits have a lower
JS than the reference ($\approx.$ 0.46) but they start as
JS$\approx$0.42.

These results are interesting when considering the noise levels in
Figure~\ref{mapping}. The yellow reference circuit (with results in
Figure~\ref{toffoli4_ol3}) chooses two connections with relatively
high noise and utilizes about 40 CNOTs, but qubits have relatively
high readout fidelity. Nonetheless, it performs better than than the
reference circuit in Figure~\ref{toffoli4_worst}, which has relatively
good connections and only 30 CNOTs.
%
The results indicate that CNOT error cannot be the only source of
noise influencing results.

Likewise, the blue mapping has one bad connection, but it provides the
best performing circuits (Figure~\ref{toffoli4_best}) with about the
same readout fidelity as for yellow. The worst results (Figure~\ref{toffoli4_worst}) contribute few (if any) good circuits,
yet benefit from relatively good connections but lower readout
fidelity according to IBM's noise data.

We know from Observation 6 that increasing CNOT error provides
additional opportunities for approximate circuits. Our mapping study
is an indication that other noise sources contribute as well,
particularly read-out errors (as depicted in Figure~\ref{mapping}) as
well as cross-talk (not reported by IBM but also known to be of the
same magnitude). This aspect requires further investigation.

\textbf{Observation 9: Sources other than CNOT error appear to
  contribute to the performance of approximate circuits.}

\subsection{Roadmap and Future Work}

Noisy gates enable and encourage circuit approximations. We plan to
extend this study and correlate circuit behavior with commonly
accepted hardware evaluation metrics, such as gate, read-out, and
cross-talk fidelity, and also ``quantum volume''~\cite{Bishop2017QuantumV}. This will allow us to project the potential of
approximations in the face of continuous hardware evolution and
decreasing noise. Metrics such as quantum volume capture the impact of
the relatively short chip coherence times. A ``small'' quantum volume
indicates there are empirical practical bounds on the circuit depth,
where we can expect approximations to benefit. Finally, best circuit
selection is performed using simulation/execution and examining the
result in its specific context. In order to guide circuit generation
and synthesis from first principles, we are interested in a thorough
analysis of the numerical value of different metrics (Hilbert-Schmidt distance~\cite{Gilchrist_2005}, Kullback-Leibler divergence~\cite{Kullback_1951}, Jensen-Shannon distance~\cite{endres_03}, etc.)

We are also looking into both deeper and wider circuits. QSearch
begins to require a prohibitive amount of search time when exposing it
to more than four qubits. QFast is a little faster and can typically
work with up to six qubits, but is still restricted in the number of
qubits it can handle.  The Berkeley Quantum Synthesis Kit recently
acquired another method of synthesis, QFactor~\cite{qfactor}, with the
ability to synthesize circuits of up to eight qubits. QFactor may be able
to create approximate circuits in that range, but a new method of
developing approximate circuits is needed for even wider circuits.

One possible solution to consider is that of breaking a large program
into pieces; it may be possible to create a large circuit out of many
small circuits, and we are interested in assessing if approximate
circuits also prove to be useful in such a context.

\section{Related Work}

While finding an approximate circuit is typically seen as less
desirable than finding an exact circuit, much work has been put in in
an effort to finding approximate circuits. The Solovay-Kitaev
algorithm \cite{SK2005} is well known to generate quantum gates which
have a specified accuracy.

Work on circuit synthesis \cite{qs,qfast,Amy2013,DeVos2015} is
often classified as either ``exact'' or ``approximate''.  But even the
approximate algorithms often end up finding closer approximations for
circuits than those we are interested in; the small allowable error
does not add enough wiggle room to take advantage of short
circuits. We are most interested in $\epsilon$-approximate synthesis
techniques, which can be coarsened to find circuits which are ``more
approximate''. Closely related is the Quantum Fast Circuit Optimizer (QFactor)~\cite{qfactor}, a newly developed piece of synthesis
software being distributed as part of the Berkeley Quantum Synthesis
Toolkit, just as QSearch and QFast are.
It can handle a greater number of qubits than QSearch and QFast can, but is focused more on circuit optimization than just synthesis, and works through tensor networks.

The Quantum Approximate Optimization Algorithm  (QAOA)~\cite{farhi2015quantum} can also be said to
create approximate circuits, though it differs from the work done here in that 
there is not a known target circuit. Much work has gone into optimizing QAOA circuits.
Especially interesting with relation to this work is the work of \cite{alam20}, which
reorders gates in order to reduce circuit length, similarly finding that approximate 
circuits with fewer CNOTs tend to outperform approximate circuits with more.

Much more work is being done on other ways to reduce noise or
circuit depth \cite{ding20,Murali2019,li2019tackling,wilson20,gokhale20,ibm-pulses,wille2016look}.
We are optimistic about these being able to work alongside
approximate circuits, though it is unclear whether the benefits
of approximate circuits will hold for process which require
post-processing or manipulation of error levels, as these may
end up interfering with the noise which the approximate circuits
rely on to perform better than exact circuits.

\section{Conclusion}

Experimental results confirm that on NISQ devices approximate circuits
have the potential to outperform theoretically precise reference
circuits. Even though these circuits would perform worse on a perfect
machine, if they are created to be similar to the reference circuits
but have fewer CNOT gates, these approximate circuits produce higher
fidelity results.

Because these improvements rely on reducing the number of CNOT gates,
we see approximate circuits perform best relative to reference
circuits in situations where the reference circuit has many CNOT
gates, namely by up to 60\% in experiments.
%

We have shown that approximate circuits can show greatly increased
performance, but we have also shown that selecting the proper
approximate circuit is more complicated than comparing process
metrics. At the very least, target machine noise levels need to be
taken into account.  Finding a reliable way to determine the ideal
approximate circuit remains an open problem.

\bibliographystyle{IEEEtran}
\bibliography{bibliography,mybib}


\end{document}